\documentclass[a4paper,twocolumn]{article}

\usepackage{graphicx}
\usepackage{upgreek}
\usepackage{authblk}
\usepackage[labelfont={bf}]{caption}
\usepackage{fancyhdr}
\usepackage{microtype}
\usepackage{balance}
\usepackage{cite}
\usepackage{hyperref}
\hypersetup{pdfborder=0 0 0}

\title{\textbf{Streamers in air splitting into three branches}}

\author[1]{L. C. J. Heijmans}
\author[1]{S. Nijdam}
\author[1]{E. M. van Veldhuizen}
\author[1,2]{U. Ebert}

\affil[1]{Department of Applied Physics, Eindhoven University of Technology - P.O. Box 513, 5600 MB Eindhoven, The Netherlands}
\affil[2]{Centrum Wiskunde and Informatica (CWI) - P.O. Box 94079, 1090 GB Amsterdam, The Netherlands}

\date{July 2013}

\begin{document}

\twocolumn[
  \begin{@twocolumnfalse}
    \vspace{20mm}
    \maketitle
    \begin{abstract}
    We investigate the branching of positive streamers in air and present the first systematic investigation of splitting into more than two branches. We study discharges in 100~mbar artificial air that is exposed to voltage pulses of 10~kV applied to a needle electrode 160~mm above a grounded plate. By imaging the discharge with two cameras from three angles, we establish that about every 200th branching event is a branching into three. Branching into three occurs more frequently for the relatively thicker streamers. In fact, we find that the surface of the total streamer cross-sections before and after a branching event is roughly the same.
    \end{abstract}
    \vspace{10mm}
  \end{@twocolumnfalse}
  ]

\section*{Introduction}

When a high electric voltage is suddenly applied to ionisable matter like air, streamers occur as rapidly growing fingers of ionised matter that due to their shape and conductivity enhance the electric field at their heads. This allows them to penetrate into regions where the background field was below the breakdown value before they approached it. On their path, streamers are frequently seen to branch.

The streamers form the primary path of a discharge that can later heat up and transform into a lightning leader~\cite{Bazelyan2000,Williams2006a} or a spark~\cite{Bazelyan1998,Gallimberti2002}. Streamers are also the main ingredient of huge sprite discharges in the thin air high above thunderstorms~\cite{Pasko2007,Ebert2010}. Streamers also have important applications in initiating gas chemistry in so-called corona reactors where the later heating phase is avoided by limiting the duration of the voltage pulse~\cite{Veldhuizen2000,Fridman2005}.

The streamer head consists of an ionisation wave that moves with velocities ranging from comparable to the local electron drift velocity to orders of magnitude faster. On these time scales, the energy is in the electrons and then in the excited and ionised atoms and molecules in the gas, while the background gas initially stays cold. This is the reason why this process is used for very energy efficient gas chemistry, with applications like, for example, gas and water cleaning~\cite{clements1989,Veldhuizen2000,Grabowski2006,Winands2006a}, ozone generation~\cite{Veldhuizen2000}, particle charging~\cite{Veldhuizen2000,Kogelschatz2004} or flow control~\cite{Moreau2007,Starikovskii2008}. An important factor for the gas treatment is which volume fraction of the gas is treated by the discharge, and this fraction clearly is determined by the branching behaviour.

Another question concerns the similarity between streamers at normal pressure and sprite discharges at air pressures in the range from mbar to $\upmu$bar at 40 to 90~km altitude in the atmosphere~\cite{Ebert2010}. Recently Kanmae \emph{et~al.}~\cite{Kanmae2012} stated, citing a private communication with Ebert in 2010, that in contrast to sprite discharges, laboratory streamers typically split into two branches only. Indeed, many streamers form out of the primary inception cloud around a needle electrode~\cite{Briels2008c}, but a propagating streamer in the lab typically splits into only two branches. There are only occasional reports of splitting into three branches~\cite{BrielsThesis_on3branch}, but these events could be a misinterpretation of images that show only a two-dimensional projection of the actual three-dimensional branching event.

Theory cannot follow the full branching dynamics either. The present stage of understanding is that the streamer can run into an unstable state that occurs when the streamer radius becomes much larger than the thickness of the space charge layer around the streamer head. This state is susceptible to a Laplacian instability~\cite[and references therein]{Luque2011a}. While this instability can develop into streamer branching in a fully deterministic manner, electron density fluctuations in the leading edge of the ionisation front accelerate the branching process. However, present simulations only can determine the time and conditions of branching, but not the evolution of the branching structure after the instability.

Studies on the full electrical discharge trees are based on dielectric breakdown models~\cite{Niemeyer1984,Pasko2000,Akyuz2003}. In these studies, a fractal-like structure is assumed for the discharge tree. The appearance of branchings is included in a phenomenological manner. Development of these models would greatly benefit from thorough knowledge on the occurrence of streamer branching.

The present paper is therefore devoted to a systematic investigation of branching into three new channels in air under laboratory conditions. In the remainder of this paper, this event will be referred to as a three-branch. As positive streamers are much easier to generate and much more frequently seen, the investigation is limited to positive streamers.

\section*{Stereo photography}

Most of the previous experimental streamer investigations are  based on two-dimensional images of streamer discharges. In reality, streamers are however a three-dimensional phenomenon. In imaging a 3D phenomenon in 2D, part of the information is lost. Some details may not be visible at all, because the line of sight of the camera is obscured.

For the present study, it is important to regard this effect. When, from the point of view of the camera, two streamers are located behind each other, they cannot be individually imaged. Instead, they will overlap on the camera image. If one of these streamers splits in two branches, these two branches combined with the other, continuing, streamer will, in a 2D projection, look like a three-branch.

A stereo photography setup has been introduced by Nijdam \emph{et al.}~\cite{Nijdam2008}. This allowed simultaneous measurement of a streamer discharge from two viewing angles, different by roughly $10^\circ$. Using this it was possible to make 3D reconstructions of a streamer discharge and measure the real (3D) branching angles. This has also been used to study the reconnection and merging of streamers~\cite{Nijdam2009}.

A different setup is used by Ichiki \emph{et~al.}~\cite{Ichiki2011,Ichiki2012} to measure branching angles in atmospheric and underwater streamers. They image each discharge from three angles. For the reconstruction two images from $0^\circ$ and $90^\circ$ are used. This large angle allows for better depth resolution compared to the $10^\circ$ angle used by Nijdam \emph{et al.}. Identifying the same streamer in both views is however much more difficult. Therefore Ichiki \emph{et~al.} used an additional image from a $225^\circ$ angle to facilitate the identification of the same streamer in the two views.

In the present study, identification of the streamers is more important than an accurate depth-coordinate. Therefore a stereographic setup based on the setup employed by Nijdam \emph{et~al.} will be used. Below, we will show that imaging at a third angle is necessary for unambiguous identification. Therefore the setup is extended with an additional camera.

It should be noted, that as far as the authors are aware, no 3D reconstructions of sprite streamers are available. Producing this would be very difficult as it would require two telescopic cameras (such as for example the one used by Kanmae \emph{et~al.}~\cite{Kanmae2012}), both aimed at the right (on forehand unknown) sprite location.

\section*{Setup}

The used point-plane discharge setup is extensively described by Nijdam \emph{et al.}~\cite{Nijdam2010}. A positive voltage pulse of, in our case, 10~kV with a rise time of about 60~ns is applied to a sharp tip to initiate streamers. The streamers propagate toward a grounded plate 160~mm below the tip. The high voltage is created with the so-called C-supply, as described extensively by Briels \emph{et al.}~\cite{Briels2006}. In this setup a charged capacitor is discharged through a spark gap switch. This creates a positive voltage pulse on the tip.

The vessel is filled with 100~mbar of artificial air. This is a pre-mixed gas mixture consisting of 20\% oxygen and 80\% nitrogen, both with less than 1~ppm contamination. These conditions were chosen such that the resulting images showed a reasonable number of branches per discharge on the one hand, but on the other hand were not so crowded that individual streamers could no longer be identified within the two views. For comparison: a pressure of 100~mbar is equal to the conditions in the earth atmosphere at 16~km altitude.

\begin{figure}
  \includegraphics[width=\linewidth]{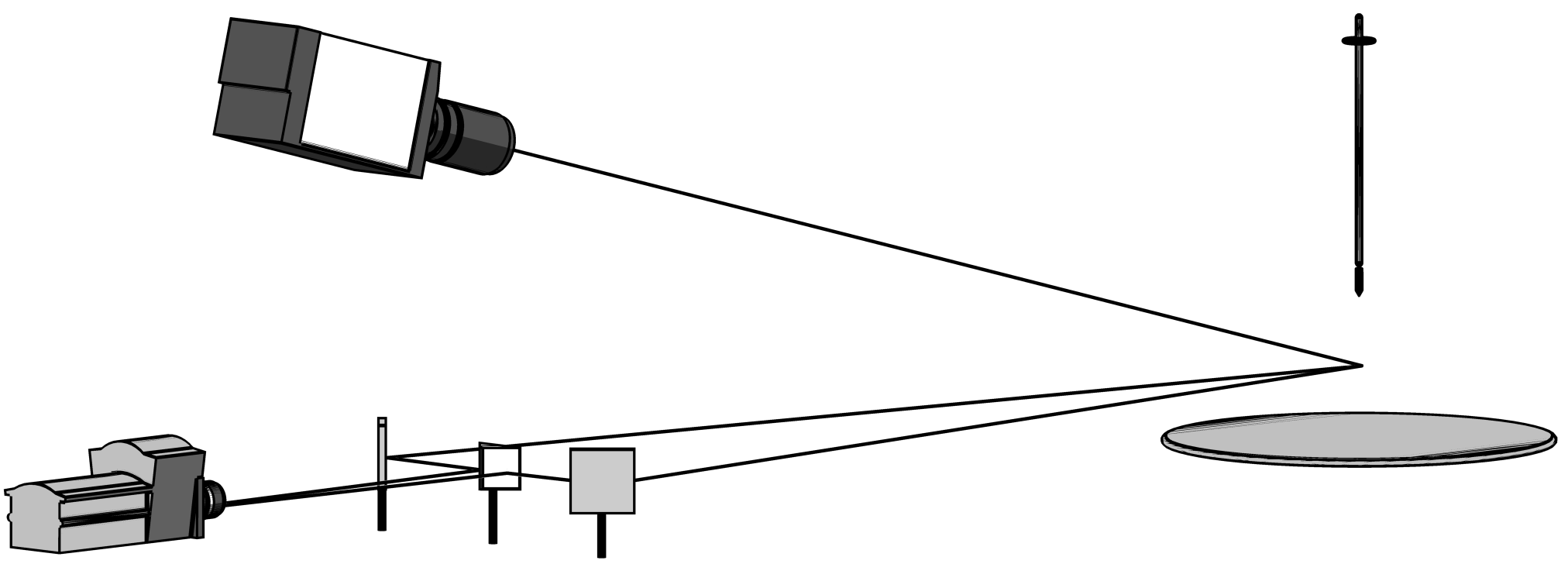}
  \caption{Simplified drawing of the used setup.The first camera is shown in the bottom left. The stereography setup, consisting of a central prism with reflecting sides and two mirrors, can be seen to its right. The second camera is visible in the top left. The point-plate discharge geometry is depicted on the right. The vacuum vessel enclosing the discharge is omitted for clarity. Lines are added to indicate the different viewing angles.}
  \label{fig:setup}
\end{figure}

The setup is schematically drawn in fig.~\ref{fig:setup}. The tip-plane geometry is depicted on the right. The surrounding vacuum vessel is omitted from the drawing. Two cameras are shown in the top and the bottom left corner. The bottom camera images the streamers through a stereographic setup as explained by Nijdam \emph{et~al.}~\cite{Nijdam2008,Nijdam2009}. Lines are added, indicating the different angles at which the streamers are imaged.

\begin{figure}
  \includegraphics[width=\linewidth]{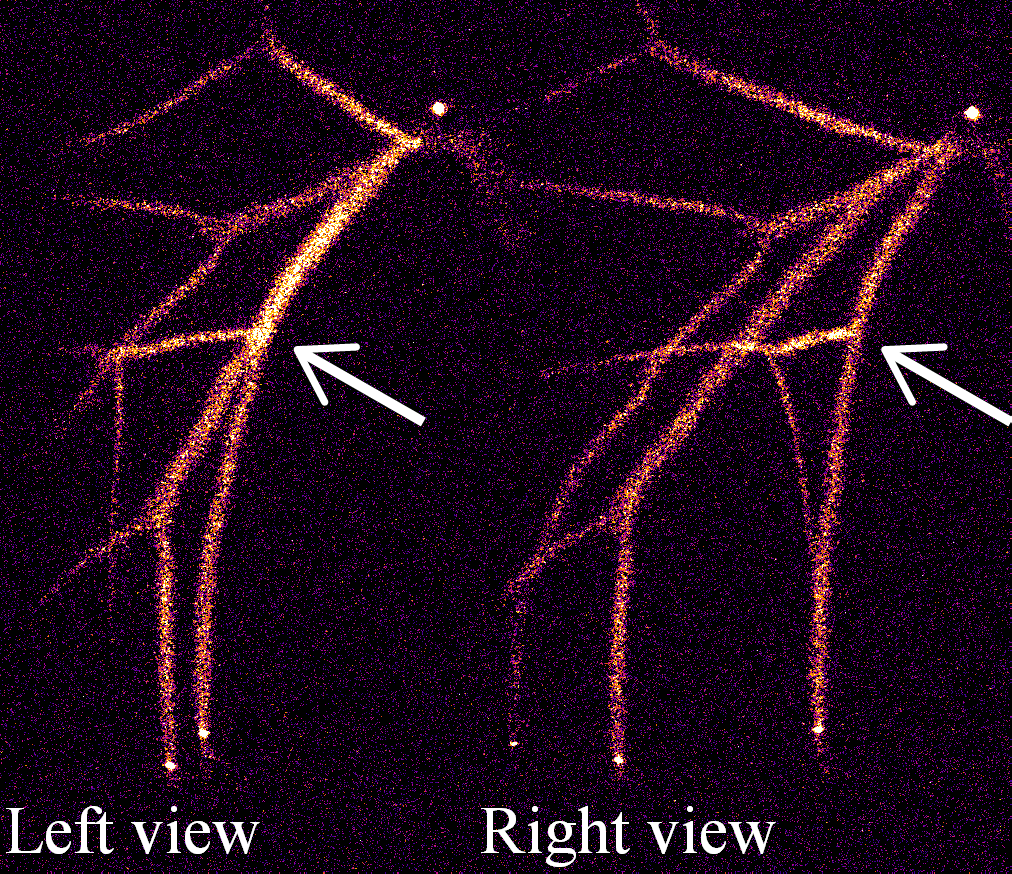}
  \caption{Example of a discharge. The discharge is imaged through a stereo photographic setup. Therefore it is shown twice; once slightly from the left, once slightly from the right. A branching event that is explained in the text is indicated with an arrow.}
  \label{fig:example}
\end{figure}

An example image of a discharge as imaged by the bottom camera using the stereo photography setup is shown in fig.~\ref{fig:example}. It shows the discharge twice; once with a viewing angle slightly from the left and once slightly from the right. In both views the tip is in the top-right corner. Only the left half of the discharge is imaged.

One branching event is indicated with a white arrow. In the left view this branching appears to be a three-branch. When looking at the right view, it is however clear that in reality it is a streamer splitting in two with a second streamer propagating in front of it. In only a single 2D image, this two-branch would have been mistaken for a three-branch.

It has been noticed that even when using the stereo photography setup, it is still possible that two streamers coincide from both viewing angles. This happens if they propagate closely behind each other, especially when they propagate (almost) horizontally. To circumvent this problem a second camera was placed in the setup. It was positioned above the original camera and images the streamers in a downward direction, as depicted in fig.~\ref{fig:setup}. In the final configuration there is a horizontal angle of $12^\circ$ between the left and the right view and a vertical angle of $15^\circ$ with the top view.

\section*{Results}

\begin{figure}
  \includegraphics[width=\linewidth]{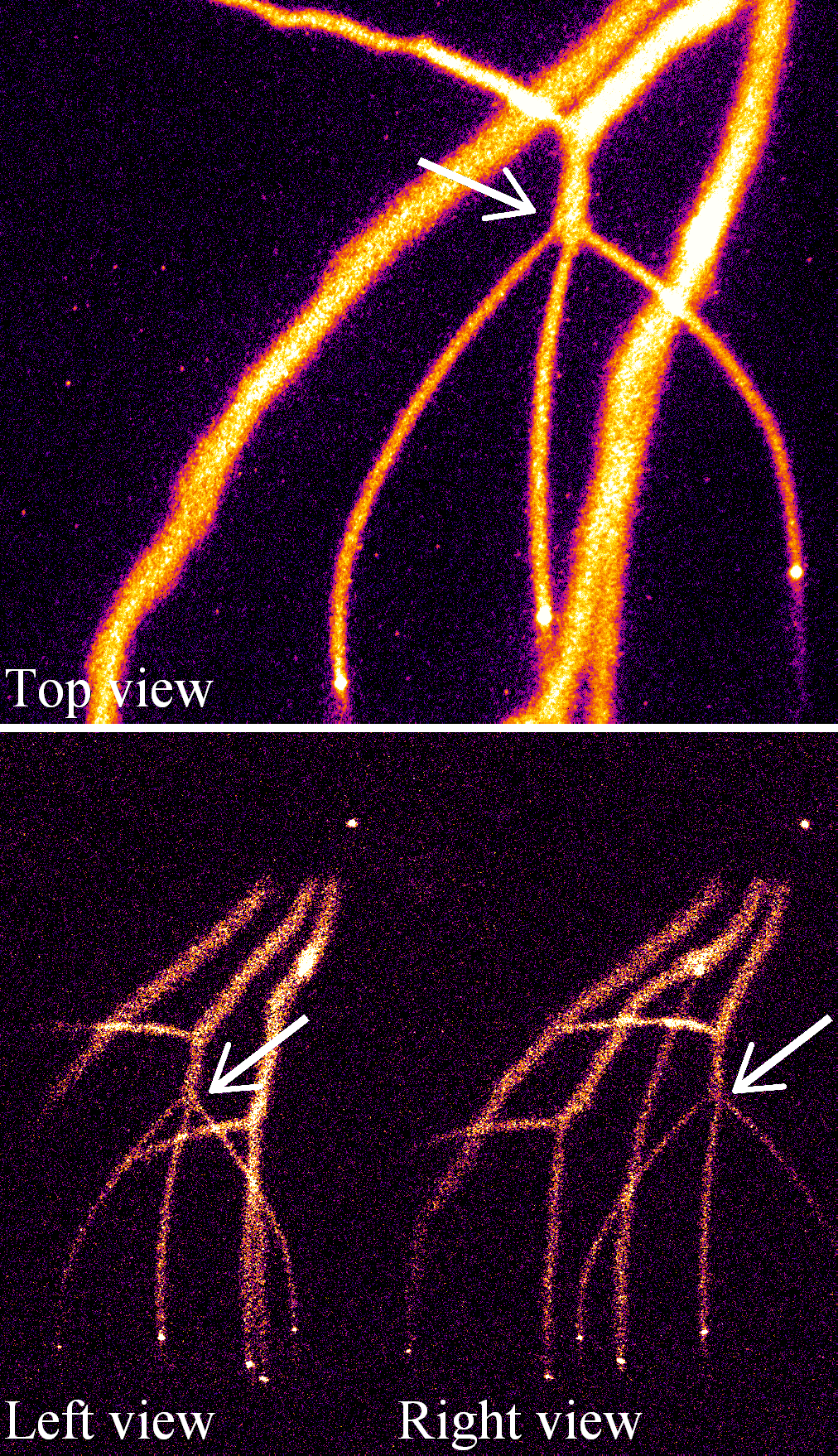}
  \caption{Example of a discharge viewed from three directions as described in the text. A branching event that is explained in the text is indicated with an arrow.}
  \label{fig:threebranch1}
\end{figure}

Figure~\ref{fig:threebranch1} shows the two images of one discharge acquired with both cameras. The top image shows the image from the top, downward looking, camera and the bottom image shows the images acquired through the stereo photography setup showing the left and the right view.

The branching event indicated with the arrow in the figure is a three-branch. This is visible from all three viewing angles. This confirms the existence of three-branches in laboratory experiments.

2187 discharges in total have been imaged. From these images, a total of 18 three-branches have been identified. On average $1.6 \pm 1.3$ branching events per picture can be identified in all three views, where the indicated error is the standard deviation of counting the number of visible branchings in 98 pictures. We estimate that roughly one out of 200 branching events under the used conditions is a three-branch. It should however be noted that linking the branches in the different views is a manual task and estimating the possible identifiability of three-branches from an image is highly tedious.

In the present study only one set of conditions (10~kV pulses in 100~mbar artificial air) is studied. Therefore no conclusion on the influence of different conditions on the number of three-branches can be drawn.

No events of streamers branching in four or more branches have been observed in the present study. If they exist, they are obviously more rare than three-branches under the given conditions.

\section*{Branching distance}

A three-branch can also be interpreted as a streamer forming a two-branch twice within a short propagation distance. If the propagation between these two subsequent branching events is small (order of the streamer thickness) it is reported as a three-branch. Measurements on the distribution of the distance between subsequent branchings would indicate whether the branching in three is a special case or that it is just an extreme in the tail of this distance distribution.

\begin{figure}
  \includegraphics[width=\linewidth]{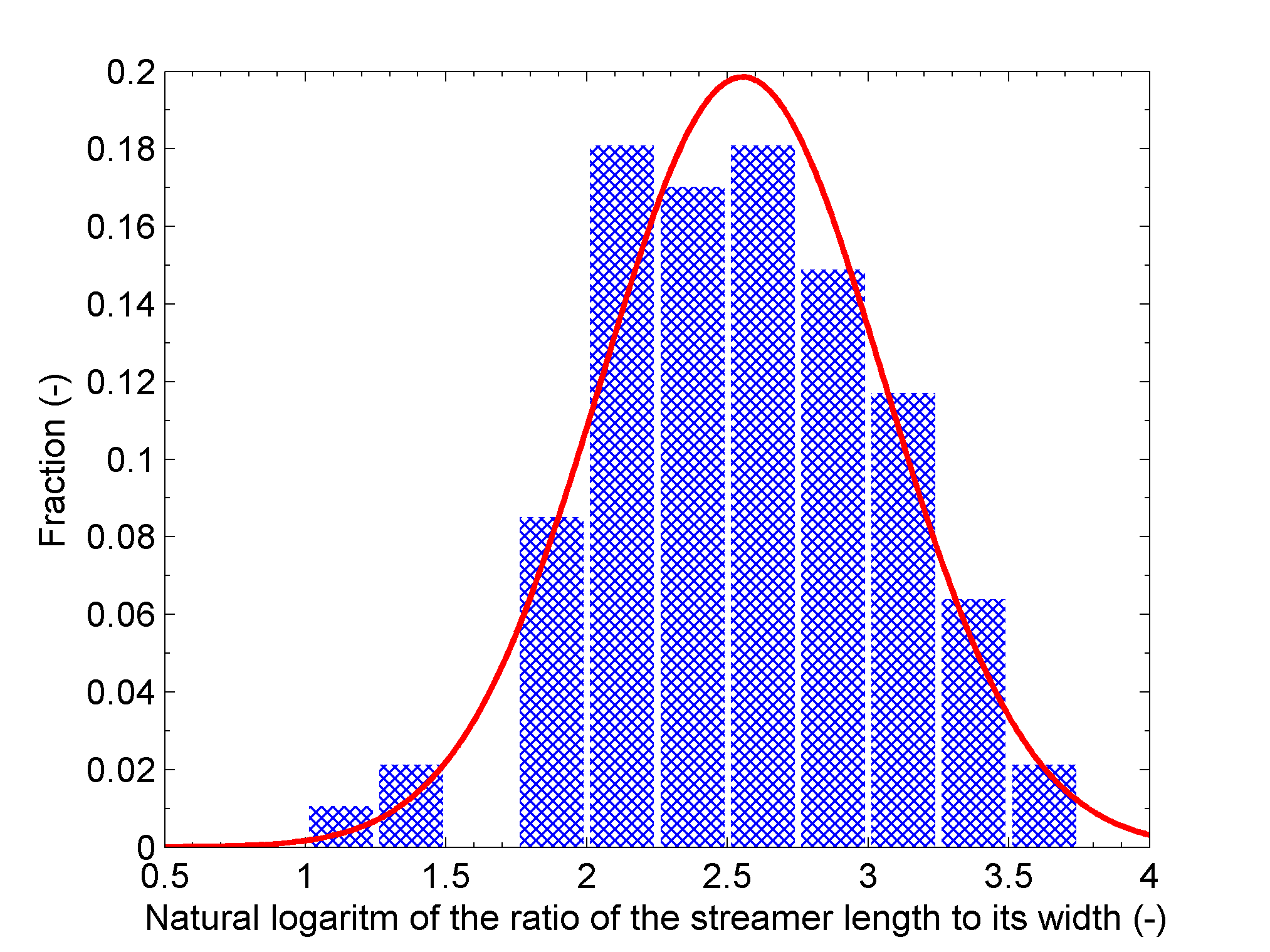}
  \caption{Normalised histogram of the natural logarithm of the ratio of the streamer length between two branchings and its width as measured by Nijdam \emph{et~al.}~\cite{Nijdam2008} (blue hatched bars) with a Gaussian distribution fit (red line).}
  \label{fig:stream_length_to_widths_ratio}
\end{figure}

For this comparison, the data measured by Nijdam \emph{et~al.}~\cite{Nijdam2008} have been analysed. They measured the ratio between the streamer length between two branchings and its width for 94 streamers.  This was done for discharges with 47~kV pulses in a 14~cm point-plane gap geometry filled with 200, 565 or 1000~mbar of ambient air. Figure~\ref{fig:stream_length_to_widths_ratio} shows a histogram of the natural logarithm $\ln(L/d)$  of this ratio. Note that in this figure data for all three pressures have been combined, as no significant difference in the ratio was found for the different pressures.

A Gaussian distribution has been fitted through the ratio distribution, as shown in fig.~\ref{fig:stream_length_to_widths_ratio}. It should be noted that the choice of fitting a Gaussian distribution on the logarithm of the ratio is purely based on the visible shape of the shown histogram and not on a physical theory regarding the expected distribution.

From the Gaussian fit it has been calculated that there is a chance of 1:1000 that a streamer would branch twice within propagating its own width (i.e. ratio~$\leq$~1). This is less often than the number of three-branches of one out of $200$ reported above.

It should be noted that the ratio measurements performed by Nijdam \emph{et~al.} were conducted under different conditions than the present measurements, namely in a smaller gap with a higher applied voltage at higher pressures in slightly different gas (ambient air versus artificial air). The ratio measurements however did not appear to depend on the gas pressure.

Secondly it should be noted, that the available data set is limited. Few data points are available in the low ratio region, therefore a large error is introduced in the extrapolation of the fitted distribution. Taking the extremes in the 95\% certainty interval for the fitted parameters of the Gaussian distribution, the chance of a ratio~$\leq$~1 can range from one in 1:100 to 1:10\,000.

As explained above, the choice for a Gaussian distribution is arbitrary and has no physical basis. Therefore the range given above can be even larger when assuming other distributions. Further measurements on the distance between subsequent streamer branchings are thus desirable.

\section*{Streamer widths}

\begin{figure}
  \includegraphics[width=\linewidth]{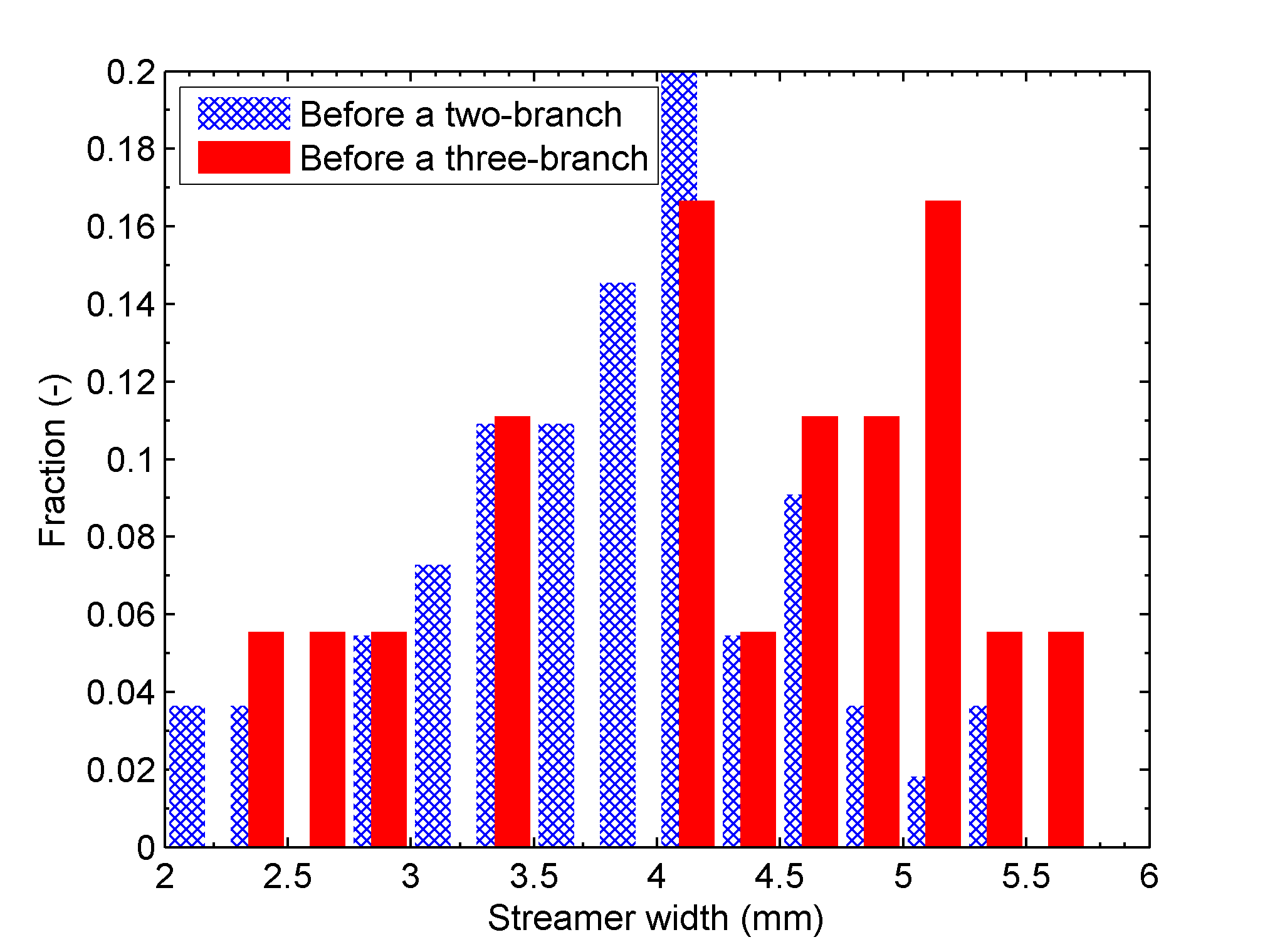}
  \caption{Normalised histogram of the width of a streamer before a two-branch (blue hatched bars) and a three-branch (red solid bars).}
  \label{fig:threebranch_widths_before}
\end{figure}

Figure~\ref{fig:threebranch_widths_before} shows a histogram of the widths of the 18~streamers, just before a three-branch. The widths have been determined in the same manner as explained by Nijdam \emph{et~al.}~\cite{Nijdam2010}. The streamer width is determined as the full width at half maximum of the average of multiple cross sections along the streamer channel. Note that the widths shown in the figure  are the average of the widths measured in the left and the right view of the discharge.

Beside the three-branches, many other (two-)branches are seen in the imaged discharges. For comparison a normalised histogram of the width of 55~streamers before such a two-branch is also displayed in the figure. It can be seen that relatively thick streamers are more likely to form a three-branch than thinner streamers. The average thickness of a streamer before a two-branch is $3.8 \pm 0.8$~mm, whereas the average thickness before a three-branch is $4.3 \pm 1.0$~mm. The given uncertainties are the standard deviation of the width distribution. Note that according to Student's t-test the chance of these width distributions being from populations with equal means is 4.7\%. This significance level is limited by the low number of measured three-branches.

Streamers at higher reduced electric fields are generally thicker. Therefore different conditions may lead to more three-branches. This might explain why streamers splitting in more than two branches are observed more often in sprite streamers than in laboratory experiments, as their reduced diameter is larger~\cite{Kanmae2012}.

Note that the width a streamer appears to have in an image is dependent on the distance to the camera. Especially if the streamer is not in the focal plane, it will appear wider than it really is. The high voltage electrode is in the focal plane of the camera. However, as the streamer discharge is three dimensional, some streamers will propagate in front or behind this focal plane. Therefore the reported diameters are an upper limit for the real streamer diameters. As no dependence of the appearance of two or three-branches on the position has been found, this effect will be equally large for two and three-branches. Therefore comparison between the two is valid even though widths are somewhat overestimated.

\begin{figure}
  \includegraphics[width=\linewidth]{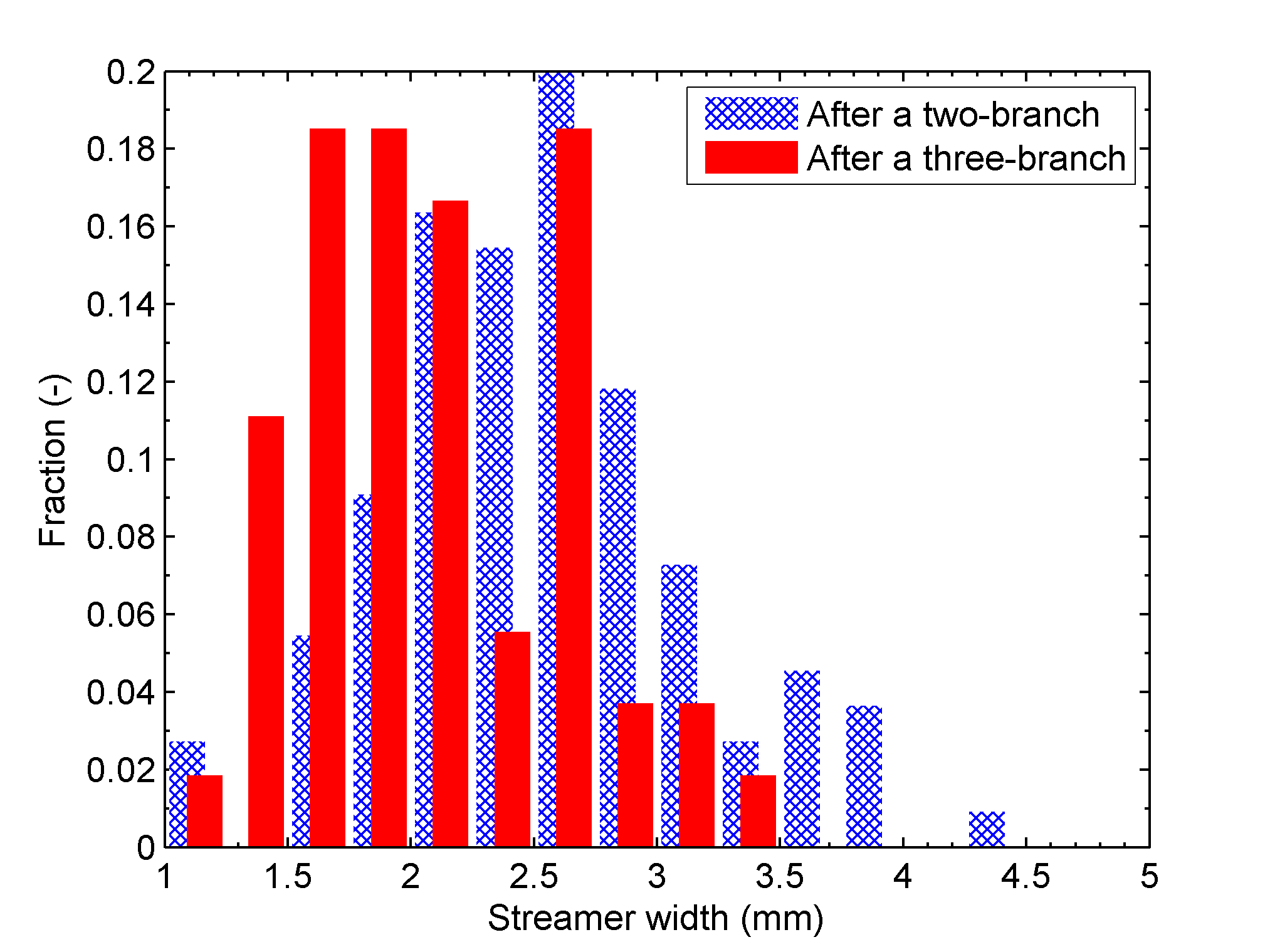}
  \caption{Normalised histogram of the width of the streamer branches after a two-branch (blue hatched bars) and a three-branch (red solid bars).}
  \label{fig:threebranch_widths_after}
\end{figure}

Figure~\ref{fig:threebranch_widths_after} shows a histogram of the widths of streamer branches after a two and a three-branch. This data is obtained from the same branching events as the data in fig.~\ref{fig:threebranch_widths_before}, but now for the two or three streamers after the branch. It is immediately clear that these branches are on average thinner than the streamers before the branching event; $2.5 \pm 0.6$ and $2.1 \pm 0.5$ after respectively a two and a three-branch. According to the t-test, the p-values for the null hypothesis are less than 1\%.

Above, it was shown that the streamers before a three-branch are on average thicker than before a two-branch. This last figure however indicates that streamers after a three-branch are thinner than after a two-branch. This means that the three-branch reduces the streamer diameter more than a two-branch.

\begin{figure}
  \includegraphics[width=\linewidth]{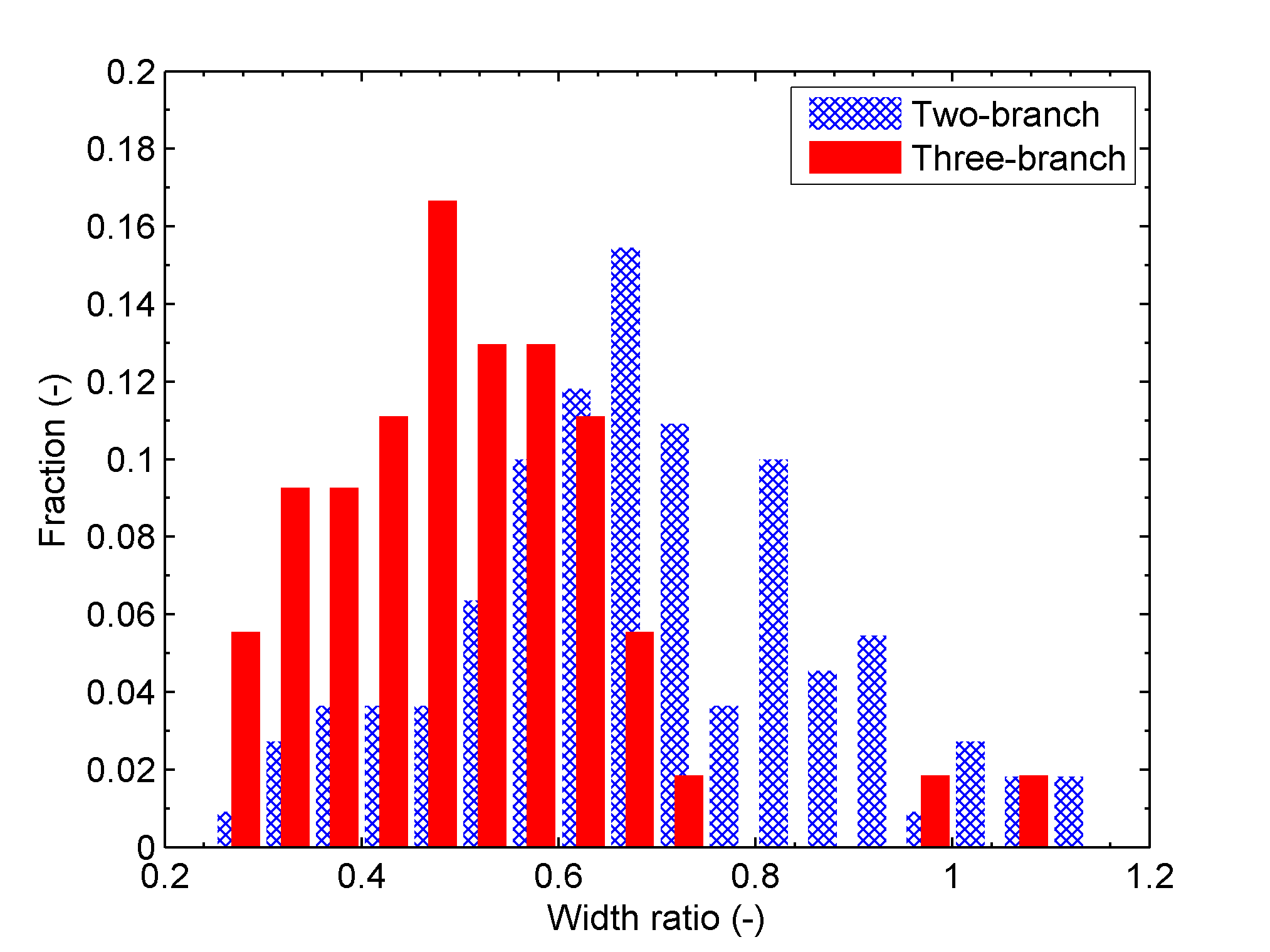}
  \caption{Normalised histogram of the ratio of the streamer widths after to before a two-branch (blue hatched bars) and a three-branch (red solid bars).}
  \label{fig:threebranch_widths_ratio}
\end{figure}

This effect is shown in more detail in fig.~\ref{fig:threebranch_widths_ratio}. Here, the ratio between the width of a streamer after a branch to its width before the branch is shown for two and three-branches. For a two-branch this ratio is $0.68 \pm 0.18$ and for a three-branch it is $0.51 \pm 0.15$. This ratio is thus smaller for three-branches, meaning these branches result in relatively thinner streamers. With the t-test, the p-value for the null hypothesis is found to be less than 1\%.

If one were to assume that the surface of the total cross section of the streamers before and after the branching would be constant, the width ratio would be~$\sqrt{1/2} \approx 0.71$ and $\sqrt{1/3} \approx 0.58$ for respectively a two and a three-branch. These values are comparable to the $0.68 \pm 0.18$ and $0.51 \pm 0.15$ found in fig.~\ref{fig:threebranch_widths_ratio}. This indicates that indeed the surface of the total cross section of the streamers before and after the branching remains approximately constant.

As a theoretical consideration: if the maximal electric field at the tip of the streamers is the same, even if they have different diameters before and after branching, the surface charge density, determining the difference between the electric field inside and ahead of the streamer, is approximately equal. If then the electric charge of the streamer is mainly concentrated at the streamer tip and the total charge would be conserved, the cross sections of parent and daughter streamers are roughly the same.

\section*{Conclusion}

It has been shown that streamer branching in three does occur in laboratory discharges created with 10~kV pulses in a 160~mm point-plane geometry filled with 100~mbar of artificial air. More than two viewing angles are required for this assessment. Under the investigated conditions it only occurs in roughly one out of $200$~branching events. This was compared to the expectation from the distance between subsequent branchings. Not enough data on the statistical distribution of this length was available for a decisive conclusion whether a three-branch is a special case or just the lower limit of the distance between two branchings.

It is shown that the three-branches on average occur in thicker streamers compared to two-branch. This might explain why it is observed more often in sprite discharges.

Streamer branches are thinner than their parent streamer both after a two and a three-branch. The reduction in diameter is bigger over a three-branch than over a two-branch. The ratio between streamers before and after a branching coincides with the value determined assuming a constant total streamer cross section surface.

\bibliographystyle{unsrt}
\balance
\bibliography{Branching_in_three_bibliography}

\end{document}